\documentclass[11pt]{article}

\usepackage[margin=1.1in]{geometry}
\usepackage{amsmath,amssymb,amsthm}
\usepackage{graphicx}
\usepackage{booktabs}
\usepackage{microtype}
\usepackage{etoolbox}
\AtBeginEnvironment{table}{\small\setlength{\tabcolsep}{4pt}}
\usepackage[]{natbib}   
\usepackage[colorlinks=true,linkcolor=blue,citecolor=blue,urlcolor=blue]{hyperref}

\graphicspath{{figures/}}

\newcommand{\Tn}{T_n}
\newcommand{\TF}{T(F)}
\newcommand{\Root}{R_n}

\title{\bf Amortized Inference for Sampling Distributions\\
Where the Bootstrap Fails}

\author{Akash Deep\\
Texas Tech University\\
\texttt{akash.deep@ttu.edu}}

\date{Draft: July 2026}

\begin{document}
\maketitle

\begin{abstract}
Efron's bootstrap is the default tool for estimating the sampling
distribution of a statistic, yet it is provably inconsistent for maxima
of bounded-support distributions, means under infinite variance, extreme
quantiles, and tail-index estimators. The classical remedies, the
$m$-out-of-$n$ bootstrap and subsampling, require rate corrections
that depend on unknown parameters and behave erratically at realistic
sample sizes. We propose an amortized alternative: a neural network is
trained on simulated datasets drawn from a prior over a distribution
family, using single independent draws of the root $\Tn - \TF$ scored by
the pinball loss, a proper scoring rule whose population minimizer is the
posterior-predictive law of the root. We call the resulting procedure a
\emph{root network}. At test time, a single forward pass
maps one dataset of $n = 200$ observations to its full
sampling-distribution estimate, from which confidence intervals follow
directly. We evaluate the method on four canonical bootstrap-failure
problems: the bounded-support maximum, the $\alpha$-stable mean, the
Pareto tail index (Hill estimator), and 99\% value-at-risk under normal
tempered stable returns. The method attains nominal 95\% coverage
(0.947 to 0.952) on all four problems, beats every feasible classical
method in Wasserstein distance to the true sampling distribution, and
captures over 97\% of the achievable improvement on the two problems
where the exact Bayes-optimal answer is computable. For the
value-at-risk problem no distribution-free method can reach nominal
coverage at all ($n=200$, $p = 0.99$ caps exact order-statistic
intervals at 85.1\%); the learned method attains 94.7\%. A single
universal network with a statistic token matches all four specialists.
On real daily market returns, evaluated against known population
quantiles of five full return series, the unchanged model averages
0.87 coverage against 0.73 for the bootstrap, matching the value
predicted by our out-of-family analysis. Stress tests delineate the
method's scope: transfer succeeds when the limit-law family is shared,
and it fails loudly and diagnosably when the convergence rate itself
changes.
\end{abstract}

\noindent\textbf{Keywords:} bootstrap inconsistency; confidence
intervals; coverage; amortized inference; simulation-based inference;
neural networks; quantile regression; heavy tails; extreme value
statistics; Hill estimator; value-at-risk; subsampling;
$m$-out-of-$n$ bootstrap; prior-data fitted networks.

\section{Introduction}
\label{sec:intro}

Given data $X_1, \dots, X_n \sim F$ and a statistic $\Tn$, frequentist
inference for the functional $\TF$ rests on the sampling distribution of
the root $\Root = \Tn - \TF$. The bootstrap \citep{efron1979} estimates
this distribution by resampling, and for smooth functionals it does so
remarkably well. But the bootstrap's failures are equally classical:
it is inconsistent for the maximum of a bounded-support distribution
\citep{bickel1981}, for the sample mean in the infinite-variance stable
domain of attraction \citep{athreya1987}, and it degrades badly for
extreme quantiles and tail-index estimators at practical sample sizes.
The standard remedies, the $m$-out-of-$n$ bootstrap
\citep{bickel1997} and subsampling \citep{politis1994}, restore
consistency in theory. In practice they require a rate correction
$(m/n)^{1-1/\alpha}$ (or its analog) that depends on \emph{unknown}
parameters. Estimating the rate \citep{bertail1999} injects noise
exactly where the method is most fragile, and the choice of $m$
remains delicate \citep{bickelsakov2008}. Our experiments quantify how
costly this is at $n = 200$: with the \emph{oracle} rate, $m$-out-of-$n$
intervals for a stable mean overcover at 99.1\% with intervals 52\%
longer than necessary, and for a 99\% quantile no distribution-free
interval can reach nominal coverage at all.

We take a different route. Specifying a prior over a family of
data-generating processes demands exactly the knowledge implicitly
invoked when choosing a resampling scheme and its rate correction.
Given such a prior, the sampling distribution of the root becomes a
\emph{learnable} object. We
simulate (dataset, root-draw) pairs from the prior and train a monotone
quantile network to predict the conditional law of the root given the
dataset; we call the trained procedure a \emph{root network}. Three
design choices make this work and are, we believe, of
independent interest:

\begin{enumerate}
\item \textbf{Single-draw proper scoring.} Each training example uses
\emph{one} independent draw of the root, scored with the pinball loss across
199 quantile levels. No Monte Carlo target distributions are ever
constructed; the proper-scoring property makes the population minimizer
the posterior-predictive law of the root given the data. This is
roughly $10^3\times$ more simulation-efficient than fitting to
empirical target distributions and structurally prevents the degenerate
solution in which a model memorizes a shared target
(Section~\ref{sec:diagnostics}).
\item \textbf{Invariance-aware representation.} The network sees
sorted, median/IQR-standardized order statistics (a sufficient
reduction of an exchangeable sample), predicts the root in
scale-equivariant units, and heavy-tailed families are handled by an
$\operatorname{asinh}$ transform of both inputs and targets, which is
legitimate because quantiles commute with monotone maps. Roots that
span many orders of magnitude across the prior (the bounded-max family
with unknown endpoint contact order) are handled by predicting in
$\log(-\Root)$ space.
\item \textbf{Own-root simulation recalibration.} Post-hoc quantile
recalibration is fit on a validation split against each dataset's
\emph{own} root $\Tn(X) - \TF$, not against independent-replicate PIT
values. The distinction matters: confidence-interval coverage is a
joint statement about the prediction and the own root, which are
dependent through $X$. Predictive-PIT recalibration left our stable-mean
coverage at 96.2\%; own-root recalibration (a learned analog of
simulation calibration, cf.~\citealp{loh1987}) achieves 95.2\%.
\end{enumerate}

\textbf{Contributions.}
(i)~A method for amortized estimation of frequentist sampling
distributions targeted at bootstrap-inconsistent statistics, with
confidence intervals as the primary output.
(ii)~An evaluation protocol anchored by \emph{provable optima}: on two
of our four families the exact Bayes-optimal data-conditional answer is
computable, so we report regret to the oracle rather than only
improvements over baselines.
(iii)~A benchmark of four canonical failure problems with exact or
high-fidelity ground truth, exact (non-resampled) implementations of the
classical baselines, and a width-tracking diagnostic that makes
input-blind models fail loudly.
(iv)~Evidence that a single universal root network with a statistic
token matches per-problem specialists, and an out-of-family analysis
that delineates scope: transfer within a shared limit-law family, loud
failure when the convergence rate changes.
Code, pretrained models, and a three-line interface for applying them
accompany the paper.

\section{Related work}
\label{sec:related}

\textbf{Prior-data fitted networks and amortized Bayes.} Our training
scheme (simulate from a prior, supervise with single draws) is the
frequentist sibling of prior-data fitted networks
\citep{muller2022pfn,hollmann2023tabpfn} and of amortized Bayesian
inference \citep{radev2020bayesflow,zammit2025review}; the theory of
PFN prediction is developed in \citet{nagler2023}. The estimand differs:
those methods approximate posterior or posterior-predictive
distributions of parameters or future observations, whereas we target
the frequentist sampling law of a root for functionals where the
bootstrap fails, and we evaluate by coverage against classical
resampling remedies. Notably, the recommended uncertainty
quantification for neural amortized point estimators is itself the
bootstrap \citep{sainsbury2024}, which is precisely what breaks in our
regimes.

\textbf{Simulation-based frequentist inference.} The LF2I family
\citep{dalmasso2021,masserano2023} obtains finite-sample-valid
confidence sets by amortizing Neyman test inversion over a parametric
simulator, and \citet{alkadhim2023} amortize p-value functions.
These methods invert tests for the parameters of a known simulator. We
instead estimate the full sampling distribution of a statistic's root,
including semi- and nonparametric functionals such as the Hill
estimator and sample maxima, and we benchmark against resampling
methods at fixed $n$. The overconfidence of amortized
posteriors under frequentist evaluation \citep{hermans2022} motivates
our coverage-first protocol.

\textbf{``Neural bootstraps.''} Several works amortize bootstrap
\emph{computation}: generators that map resampling weights to
bootstrapped predictions or estimator values
\citep{nalisnick2017,shin2021neuboots,shin2020gms,nie2022}. Their
estimand is the classical bootstrap distribution itself, so bootstrap
inconsistency is inherited, not corrected. We are aware of no prior
method that learns to \emph{replace} the bootstrap where it is
inconsistent.

\textbf{Classical remedies.} $m$-out-of-$n$ bootstrap
\citep{bickel1997}, subsampling \citep{politis1994}, estimated-rate
subsampling \citep{bertail1999}, data-driven $m$ for extrema
\citep{bickelsakov2008}, and calibrated/prepivoted intervals
\citep{loh1987,beran1987} form our baseline suite and our positioning:
these methods presume either a known rate or bootstrap consistency to
begin with.

\section{Method: Learning Sampling Distributions from Simulation}
\label{sec:method}

\begin{figure}[htbp]
\centering
\includegraphics[width=\textwidth]{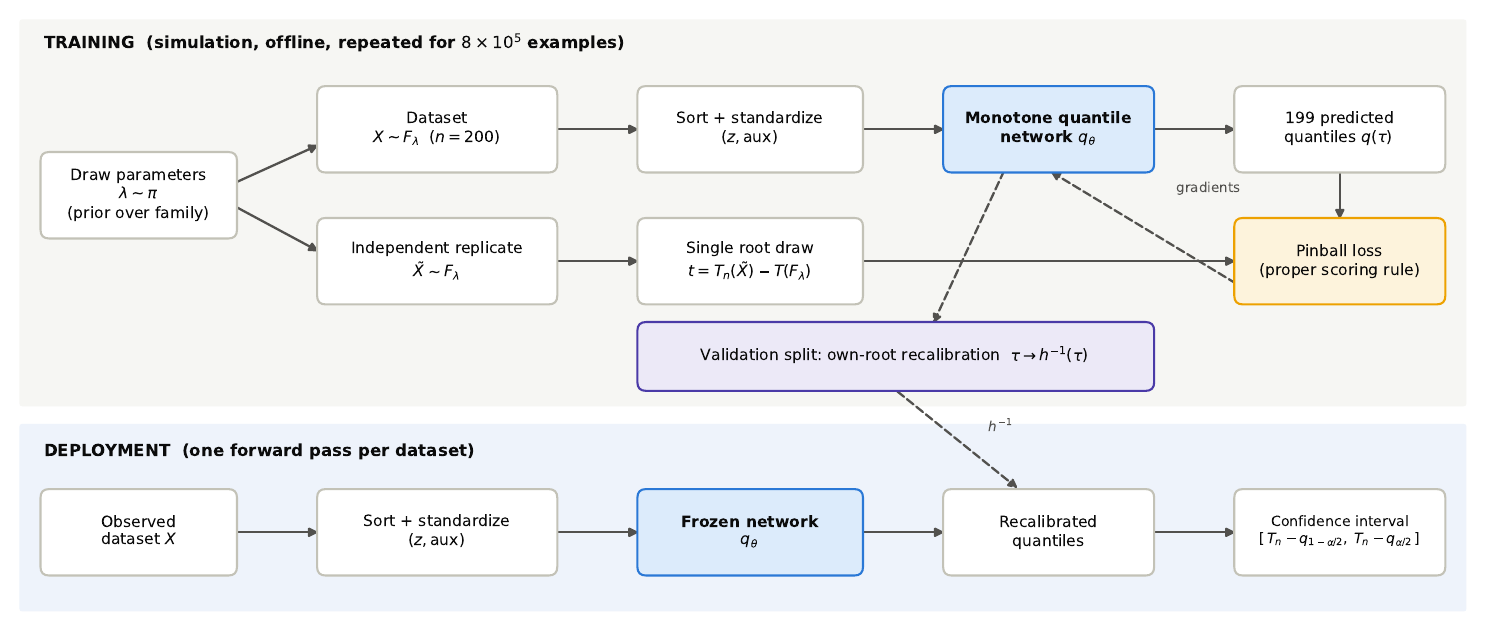}
\caption{Method overview. Training (top): parameters are drawn from the
prior; the network sees sorted, standardized order statistics of a
dataset and is scored by the pinball loss against a single root draw
from an independent replicate, so no Monte Carlo target distributions
are ever constructed; a validation split then fits the own-root
level recalibration. Deployment (bottom): a single forward pass of the
frozen network yields recalibrated quantiles of the root and the
confidence interval for $\TF$.}
\label{fig:method}
\end{figure}

Figure~\ref{fig:method} gives the complete picture: an offline training
loop that simulates (dataset, root-draw) pairs from the prior and fits
a monotone quantile network by proper scoring, followed by deployment,
where one forward pass maps an observed dataset to recalibrated
quantiles and a confidence interval. The subsections below detail each
component.

\subsection{Estimand}
Let $\pi$ be a prior over a family $\{F_\lambda\}$ of data-generating
processes, and let the observed dataset be $X = (X_1, \dots, X_n)
\stackrel{iid}{\sim} F_\lambda$ with $\lambda \sim \pi$. Our target is
the conditional law of an independent root,
\begin{equation}
\label{eq:estimand}
G(\cdot \mid X) \;=\; \mathcal{L}\!\left( \Tn(\tilde X) - T(F_\lambda)
\,\middle|\, X \right),
\qquad \tilde X \stackrel{iid}{\sim} F_\lambda \perp X,
\end{equation}
the posterior-predictive sampling distribution of the root. It is
genuinely data-conditional: a constant (input-independent) predictor is
suboptimal whenever the root law varies over the prior. Equal-tailed
intervals for $\TF$ follow from predicted root quantiles $q(\tau)$ as
$[\,\Tn - q(1-\alpha/2),\; \Tn - q(\alpha/2)\,]$.

\subsection{Single-draw training}
Each training example is generated as: $\lambda \sim \pi$; $X \sim
F_\lambda$; independently $\tilde X \sim F_\lambda$; target $t =
\Tn(\tilde X) - T(F_\lambda)$. A network $q_\theta$ maps features of
$X$ to quantiles at levels $\tau_1 < \dots < \tau_L$ ($L = 199$,
$\tau_j = j/200$) and is trained with the average pinball loss
\begin{equation}
\ell(q, t) = \frac{1}{L} \sum_{j=1}^{L}
\rho_{\tau_j}\!\left(t - q(\tau_j)\right),
\qquad
\rho_\tau(u) = u\,(\tau - \mathbf{1}\{u < 0\}),
\end{equation}
a discretized CRPS. Because $\rho_\tau$ is a proper scoring rule for
the $\tau$-quantile, the population minimizer of
$\mathbb{E}[\ell(q_\theta(X), t)]$ is exactly the quantile function of
\eqref{eq:estimand}. No empirical target distributions are simulated;
one root draw per example suffices.

\subsection{Representation and output head}
Inputs are the sorted values $z_{(1)} \le \dots \le z_{(n)}$ after
median/IQR standardization, plus auxiliary scale and location
coordinates $(\log s, \,\mathrm{med}/s)$; sorting is a canonical
sufficient reduction of an exchangeable sample. For heavy-tailed
families the standardized values and the training targets are passed
through $\operatorname{asinh}$; for the bounded-max family with unknown
contact order, targets are trained in $\log(-\Root)$ space. Both
transforms are monotone, so predicted quantiles invert exactly. The
output head enforces monotonicity by construction:
$q(\tau_1) = \beta$, $q(\tau_{j+1}) = q(\tau_j) +
\operatorname{softplus}(\eta_j)$. Unlike mixture-density heads, this
imposes no shape restrictions and represents hard-boundary roots (the
max) without leakage across the boundary. The trunk is a three-layer
MLP (width 384; $4.5\times 10^5$ parameters); a permutation-invariant
encoder is a drop-in replacement left to future work since sorted
inputs already suffice at fixed $n$.

\subsection{Own-root recalibration}
\label{sec:recal}
After training, we fit a level-recalibration map on the validation
split: with validation datasets $X_i$ (whose generating $\lambda_i$ is
known by construction), compute $h(\tau) = \tfrac{1}{m}\sum_i
\mathbf{1}\{\Tn(X_i) - T(F_{\lambda_i}) \le q_i(\tau)\}$ and replace
the level used for nominal $\tau$ by $h^{-1}(\tau)$
\citep[cf.][]{kuleshov2018}. Crucially the recalibration target is the
dataset's \emph{own} root: coverage is a joint event of the prediction
and the own root, which are dependent through $X$, and marginal PIT
calibration against independent replicates does not control it. For
location and scale families the two notions coincide by symmetry, which
is why the distinction is easy to miss; for the stable mean it is worth
a full coverage point (96.2\% vs.\ 95.2\%). Test data are never used.

\subsection{The width-tracking diagnostic}
\label{sec:diagnostics}
Amortized models with weak or misconstructed targets can converge to
input-\emph{independent} predictors that nevertheless score well
against marginal criteria. We therefore run, in every experiment, what
we call the \emph{width-tracking diagnostic}: (i) the correlation,
across test datasets, between predicted and true 95\%-interval
log-widths, and (ii) the response of predictions to pure noise inputs.
A constant predictor scores zero on both; our models score 0.95 to
1.00 on width tracking with large noise response. The diagnostic is
statistic-agnostic and costs a few forward passes; we recommend it as
a standard check for any amortized inference procedure, and it is
exposed as a first-class function in our software.
Appendix~\ref{app:cautionary} recounts the failure that motivated it.

\section{Experimental protocol}
\label{sec:protocol}

\textbf{Families.} Four canonical bootstrap-failure problems at
$n = 200$ (priors in parentheses):
uniform maximum ($\theta \sim \mathrm{LogU}(0.5, 5)$; $\Tn = X_{(n)}$,
$\TF = \theta$); symmetric $\alpha$-stable mean ($\alpha \sim
U(1.1, 1.95)$, scale $\mathrm{LogU}(0.5,5)$, location $U(-2,2)$); Hill
tail-index estimator with $k = 34$ under Pareto ($\alpha \sim U(1.5,
4)$, scale $\mathrm{LogU}(0.5,5)$); and 99\% value-at-risk under
symmetric normal tempered stable returns ($\alpha \sim U(1.1, 1.9)$,
tempering $\mathrm{LogU}(0.3, 3)$, scale, location). Training uses
$8 \times 10^5$ examples per family ($\approx$ 15 GPU-minutes each).

\textbf{Ground truth and oracles.} Uniform max, Hill, and stable-mean
root distributions are available exactly (order-statistic algebra, the
R\'enyi representation making the Hill root exactly
$\gamma(\Gamma_k/k - 1)$, and the stability property, respectively;
the standard-stable quantile table uses $2\times 10^7$ draws per
$\alpha$). NTS truth uses fresh $2 \times 10^6$-draw pools per held-out
parameter. On the uniform max and Hill families the exact
Bayes-posterior-predictive oracle under the prior is computable (a 1-D
posterior mixture in each case), giving regret-to-optimum anchors.

\textbf{Baselines.} Standard bootstrap; $m$-out-of-$n$ bootstrap and
subsampling with $m = n^{2/3}$ (naive, oracle-rate, and estimated-rate
corrections where the rate is unknown); parametric bootstraps
(exact for the max; McCulloch quantile estimation plus the stability
property for the stable mean; maximum likelihood for Pareto); the exact
order-statistic interval for quantiles; and the CLT interval. For the
max, all resampling baselines are computed \emph{exactly} from
order-statistic algebra rather than by Monte Carlo.

\textbf{Metrics.} Coverage and length of equal-tailed 95\%/90\%
intervals over $2\times 10^4$ test datasets from 100 to 200 held-out
parameters; Wasserstein-1 distance between predicted and true root
quantile functions, normalized per dataset by the true root scale.
Train/validation/test/truth randomness comes from disjoint,
index-stable seed streams. Because test datasets cluster within
parameters, we report coverage standard errors both pooled and
clustered at the parameter level (Table~\ref{tab:clustered_se});
clustering changes the SEs by at most a factor of 1.6, and every
coverage estimate remains within two clustered SEs of nominal.
\begin{table}[htbp]
\centering
\caption{95\% coverage with pooled and clustered (between-parameter) standard errors.}
\label{tab:clustered_se}
\begin{tabular}{lcccc}
\toprule
Family & Cov.\ 95 & Pooled SE & Clustered SE & Params \\
\midrule
m1\_uniform\_max & 0.949 & 0.0016 & 0.0015 & 200 \\
m2\_stable\_mean & 0.952 & 0.0015 & 0.0018 & 200 \\
m3\_pareto\_hill & 0.951 & 0.0015 & 0.0016 & 200 \\
m3\_nts\_var & 0.947 & 0.0022 & 0.0035 & 100 \\
m4c\_beta\_max & 0.955 & 0.0015 & 0.0018 & 200 \\
\bottomrule
\end{tabular}
\end{table}

\section{Results}
\label{sec:results}

\subsection{Benchmark results on four bootstrap-failure problems}
\begin{figure}[htbp]
\centering
\includegraphics[width=\textwidth]{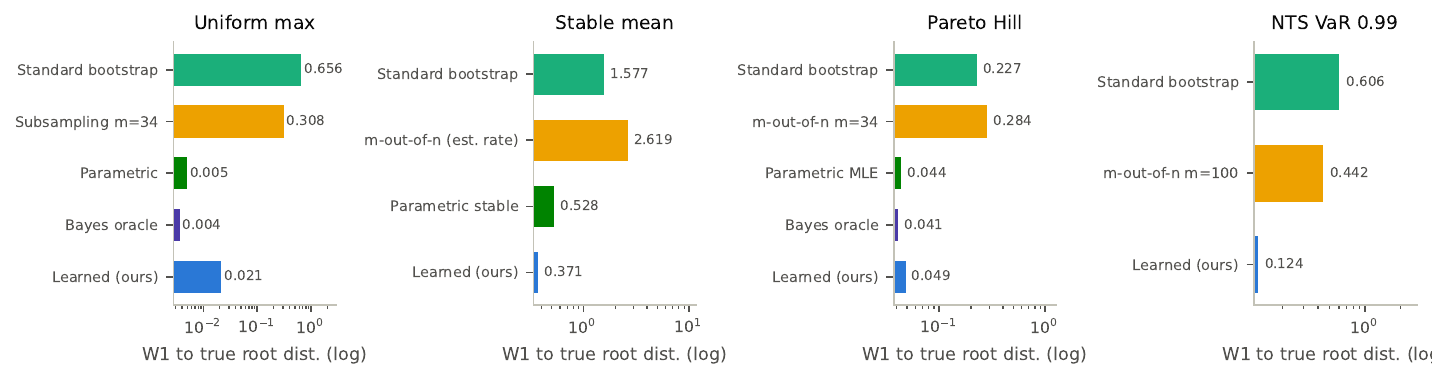}
\caption{Wasserstein-1 distance to the true root distribution by method
and family (log scale; smaller is better). The Bayes oracle rows give
the achievable optimum where it is computable.}
\label{fig:w1}
\end{figure}

Tables~\ref{tab:m1_uniform_max} through \ref{tab:m3_nts_var} report the
four families, and Figure~\ref{fig:w1} summarizes the accuracy
comparison. Three findings stand out. First, the root network is
the only method at nominal coverage on all four problems simultaneously
(0.949, 0.952, 0.951, and 0.947 at the 95\% level). Second, it beats
every feasible nonparametric method in $W_1$ by factors of 4 to 30.
Third, on the stable mean it also produces the shortest
correctly-covering intervals.

\begin{table}[htbp]
\centering
\caption{Uniform max, theta $\sim$ LogUniform(0.5, 5), n = 200. Nominal coverage is 0.95 (Cov. 95) and 0.90 (Cov. 90); interval lengths and W1 distances to the true root distribution are normalized per dataset by the true root scale. Blank cells: metric not defined for that method.}
\label{tab:m1_uniform_max}
\begin{tabular}{lccccc}
\toprule
Method & Cov. 95 & Cov. 90 & Len. 95 & W1 truth & W1 to oracle \\
\midrule
standard\_bootstrap & 0.877 & 0.750 & 3.01 & 0.656 &  \\
subsampling\_m=34 & 0.941 & 0.906 & 3.06 & 0.308 &  \\
parametric\_bootstrap & 0.949 & 0.898 & 3.61 & 0.005 &  \\
exact\_pivot\_(oracle) & 0.951 & 0.900 & 3.68 & 0.007 &  \\
bayes\_(oracle) & 0.949 & 0.898 & 3.63 & 0.004 &  \\
learned\_raw & 0.952 & 0.899 & 3.89 & 0.025 & 0.025 \\
learned\_recal\_(ours) & 0.949 & 0.901 & 3.65 & 0.021 & 0.021 \\
\bottomrule
\end{tabular}
\end{table}

\begin{table}[htbp]
\centering
\caption{Stable mean, alpha $\sim$ U(1.1, 1.95), n = 200 (Athreya case). Nominal coverage is 0.95 (Cov. 95) and 0.90 (Cov. 90); interval lengths and W1 distances to the true root distribution are normalized per dataset by the true root scale. Blank cells: metric not defined for that method.}
\label{tab:m2_stable_mean}
\begin{tabular}{lccccc}
\toprule
Method & Cov. 95 & Cov. 90 & Len. 95 & W1 truth & W1 to oracle \\
\midrule
normal\_CLT\_interval & 0.960 & 0.910 & 10.91 & 1.648 &  \\
standard\_bootstrap & 0.967 & 0.922 & 9.68 & 1.577 &  \\
m\_of\_n\_naive\_rate\_m34 & 0.975 & 0.933 & 8.84 & 1.600 &  \\
m\_of\_n\_mcculloch\_m34 & 0.989 & 0.970 & 14.35 & 2.619 &  \\
m\_of\_n\_oracle\_rate\_m34 & 0.991 & 0.973 & 13.69 & 2.402 &  \\
parametric\_stable\_(ref) & 0.934 & 0.883 & 10.41 & 0.528 &  \\
learned\_raw & 0.967 & 0.917 & 11.12 & 0.393 & 0.424 \\
learned\_recal\_(ours) & 0.952 & 0.898 & 8.99 & 0.371 & 0.419 \\
\bottomrule
\end{tabular}
\end{table}

\begin{table}[htbp]
\centering
\caption{Hill estimator, Pareto family, alpha $\sim$ U(1.5, 4), k = 34, n = 200. Nominal coverage is 0.95 (Cov. 95) and 0.90 (Cov. 90); interval lengths and W1 distances to the true root distribution are normalized per dataset by the true root scale. Blank cells: metric not defined for that method.}
\label{tab:m3_pareto_hill}
\begin{tabular}{lccccc}
\toprule
Method & Cov. 95 & Cov. 90 & Len. 95 & W1 truth & W1 to oracle \\
\midrule
standard\_bootstrap & 0.913 & 0.863 & 3.93 & 0.227 &  \\
m\_of\_n\_m34\_k10 & 0.862 & 0.800 & 3.83 & 0.284 &  \\
parametric\_MLE & 0.947 & 0.896 & 3.89 & 0.044 &  \\
bayes\_(oracle) & 0.950 & 0.900 & 3.94 & 0.041 &  \\
learned\_raw & 0.949 & 0.901 & 3.92 & 0.047 & 0.022 \\
learned\_recal\_(ours) & 0.951 & 0.903 & 3.94 & 0.049 & 0.026 \\
\bottomrule
\end{tabular}
\end{table}

\begin{table}[htbp]
\centering
\caption{VaR 0.99, NTS family, alpha $\sim$ U(1.1, 1.9), n = 200. Nominal coverage is 0.95 (Cov. 95) and 0.90 (Cov. 90); interval lengths and W1 distances to the true root distribution are normalized per dataset by the true root scale. Blank cells: metric not defined for that method.}
\label{tab:m3_nts_var}
\begin{tabular}{lcccc}
\toprule
Method & Cov. 95 & Cov. 90 & Len. 95 & W1 truth \\
\midrule
standard\_bootstrap & 0.724 & 0.683 & 4.16 & 0.606 \\
m\_of\_n\_m100 & 0.720 & 0.683 & 3.41 & 0.442 \\
binomial\_exact\_orderstat & 0.851 & 0.851 & 3.73 &  \\
learned\_raw & 0.945 & 0.892 & 4.20 & 0.112 \\
learned\_recal\_(ours) & 0.947 & 0.902 & 4.04 & 0.124 \\
\bottomrule
\end{tabular}
\end{table}

\textbf{Regret to provable optima.} On the uniform max the learned
method captures 97.3\% of the gap between the standard bootstrap and
the exact Bayes oracle; on the Hill family it sits essentially at the
oracle ($W_1$ 0.049 vs.\ 0.041 in root-scale units) while the
parametric MLE undercovers.

\textbf{Beating a parametric oracle without the family.} On the stable
mean the learned method attains $W_1 = 0.371$ against $0.528$ for a
parametric-stable bootstrap that is \emph{told} the family and uses
the exact stability property with McCulloch estimation. Two mechanisms:
the network consumes all $n$ order statistics rather than five sample
quantiles, and its predictive target integrates parameter uncertainty
that plug-in bootstraps ignore (the parametric oracle covers 93.4\%).

\textbf{Where classical methods cannot work.} For the 99\% quantile at
$n = 200$ the exact order-statistic interval has a provable two-sided
coverage ceiling of $1 - 0.99^{200} \approx 0.866$ (0.851 attained);
the standard and $m$-out-of-$n$ bootstraps cover 72.4\% and 72.0\%.
The root network covers 94.7\%.

\begin{figure}[htbp]
\centering
\includegraphics[width=\textwidth]{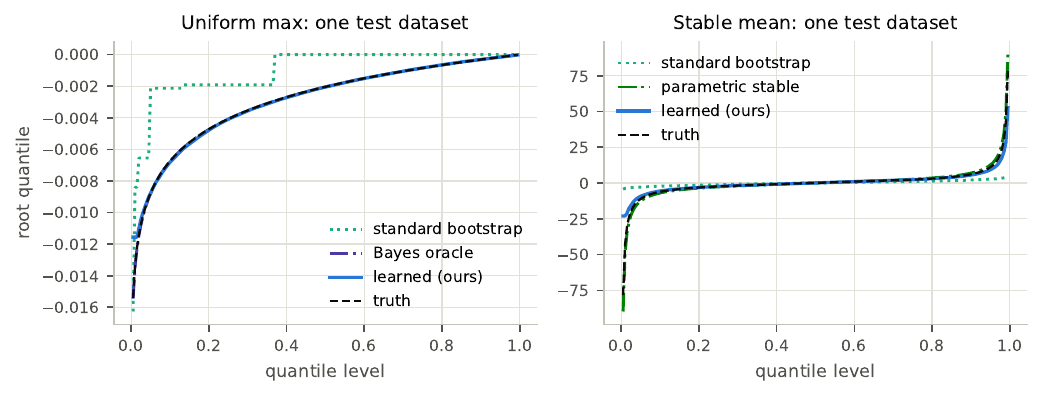}
\caption{Predicted root quantile functions for single test datasets.
Left (uniform max): the standard bootstrap is a step function with an
atom at zero of mass approaching $1 - 1/e$ (the classical
inconsistency), while the learned prediction, the exact Bayes oracle,
and the truth coincide. Right (stable mean): the bootstrap misses the
heavy tails entirely; the learned prediction tracks the truth across
the full range.}
\label{fig:overlay}
\end{figure}

\begin{figure}[htbp]
\centering
\includegraphics[width=\textwidth]{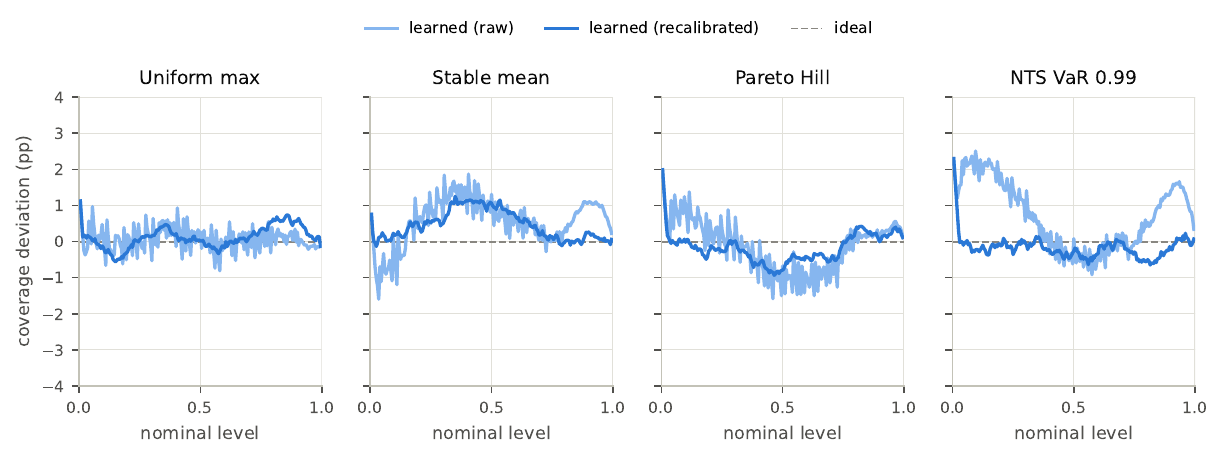}
\caption{Reliability (empirical minus nominal coverage, percentage
points) before and after own-root recalibration.}
\label{fig:reliability}
\end{figure}

\begin{figure}[htbp]
\centering
\includegraphics[width=\textwidth]{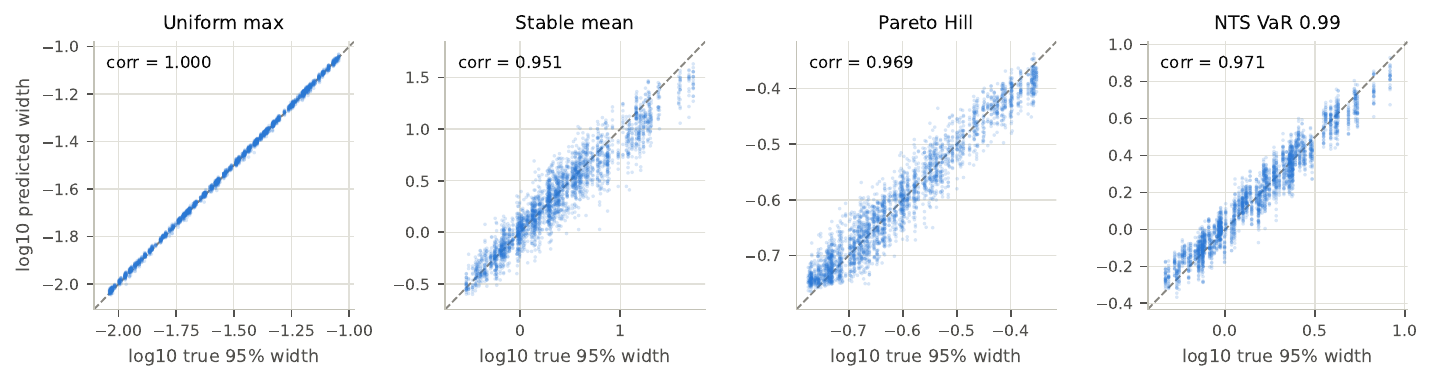}
\caption{Predicted vs.\ true 95\%-interval widths per test dataset,
on log-log axes. An input-blind model would be a horizontal line;
correlations are 0.95 to 1.00.}
\label{fig:width}
\end{figure}

\subsection{One universal network for four statistics}
A single network with a 4-dimensional statistic token, trained on all
four families at the same total budget, matches the specialists on
bit-identical test sets: coverage 0.949/0.951/0.951/0.949 vs.\
0.949/0.952/0.951/0.947, with $W_1$ within 2 to 4\%.
\begin{table}[htbp]
\centering
\caption{Universal model (statistic token) vs specialists, same test sets. Nominal coverage is 0.95 (Cov. 95) and 0.90 (Cov. 90); interval lengths and W1 distances to the true root distribution are normalized per dataset by the true root scale. Blank cells: metric not defined for that method.}
\label{tab:m4_universal}
\begin{tabular}{lcccc}
\toprule
Method & Cov. 95 & Cov. 90 & Len. 95 & W1 truth \\
\midrule
m1\_uniform\_max / specialist & 0.949 & 0.901 & 3.65 & 0.021 \\
m1\_uniform\_max / universal & 0.949 & 0.895 & 3.62 & 0.025 \\
m2\_stable\_mean / specialist & 0.952 & 0.898 & 8.99 & 0.371 \\
m2\_stable\_mean / universal & 0.951 & 0.898 & 8.69 & 0.378 \\
m3\_pareto\_hill / specialist & 0.951 & 0.903 & 3.94 & 0.049 \\
m3\_pareto\_hill / universal & 0.951 & 0.895 & 3.92 & 0.050 \\
m3\_nts\_var / specialist & 0.947 & 0.902 & 4.04 & 0.124 \\
m3\_nts\_var / universal & 0.949 & 0.901 & 4.05 & 0.128 \\
\bottomrule
\end{tabular}
\end{table}

\subsection{Out-of-family behavior and learning the rate}
\label{sec:ood}
Evaluating the specialists unchanged outside their priors yields three
regimes (Table~\ref{tab:m4_ood}): (i) when the limit-law family is
shared, transfer is conservative and still beats the bootstrap (Burr
Hill: 96.2\% vs.\ 75.4\%; Student-$t$ VaR: 86.1\% vs.\ 70.3\%;
Student-$t$ mean in the stable domain of attraction: 98.1\%); (ii) in
the regular regime the bootstrap rightly wins; (iii) when the
convergence \emph{rate} changes (endpoint contact order of the max),
coverage collapses for our method, and effectively for everything
else. The remedy is the method's own prescription: widen the prior.
\begin{table}[htbp]
\centering
\caption{Out-of-family stress tests: specialists evaluated outside their training prior. Nominal coverage is 0.95 (Cov. 95) and 0.90 (Cov. 90); interval lengths and W1 distances to the true root distribution are normalized per dataset by the true root scale. Blank cells: metric not defined for that method.}
\label{tab:m4_ood}
\begin{tabular}{lcccc}
\toprule
Method & Cov. 95 & Cov. 90 & Len. 95 & W1 truth \\
\midrule
beta\_max\_b0.5 / standard\_boot & 0.977 & 0.941 & 5.42 & 0.680 \\
beta\_max\_b0.5 / learned\_raw & 0.093 & 0.035 & 230.29 & 57.725 \\
beta\_max\_b0.5 / learned\_recal & 0.103 & 0.039 & 216.26 & 57.753 \\
beta\_max\_b2 / standard\_boot & 0.578 & 0.497 & 2.37 & 1.496 \\
beta\_max\_b2 / learned\_raw & 0.052 & 0.032 & 0.52 & 1.840 \\
beta\_max\_b2 / learned\_recal & 0.047 & 0.031 & 0.49 & 1.840 \\
t\_mean\_heavy / standard\_boot & 0.971 & 0.924 & 2.86 & 0.361 \\
t\_mean\_heavy / learned\_raw & 0.987 & 0.962 & 8.33 & 0.413 \\
t\_mean\_heavy / learned\_recal & 0.981 & 0.955 & 6.50 & 0.324 \\
t\_mean\_light / standard\_boot & 0.954 & 0.904 & 3.87 & 0.078 \\
t\_mean\_light / learned\_raw & 0.996 & 0.977 & 6.88 & 0.334 \\
t\_mean\_light / learned\_recal & 0.991 & 0.965 & 5.85 & 0.280 \\
burr\_hill / standard\_boot & 0.754 & 0.655 & 4.04 & 1.322 \\
burr\_hill / learned\_raw & 0.964 & 0.925 & 7.30 & 1.381 \\
burr\_hill / learned\_recal & 0.962 & 0.923 & 7.29 & 1.387 \\
frechet\_hill / standard\_boot & 0.927 & 0.870 & 3.96 & 0.392 \\
frechet\_hill / learned\_raw & 0.994 & 0.982 & 6.02 & 0.494 \\
frechet\_hill / learned\_recal & 0.994 & 0.981 & 6.03 & 0.501 \\
t\_var99 / standard\_boot & 0.703 & 0.670 & 4.56 & 0.659 \\
t\_var99 / learned\_raw & 0.860 & 0.782 & 3.19 & 0.181 \\
t\_var99 / learned\_recal & 0.861 & 0.798 & 3.04 & 0.192 \\
\bottomrule
\end{tabular}
\end{table}

\textbf{Learning the rate.} We test the prescription directly: retrain
the max model on $F_V(v) = 1 - (1-v)^b$ with the contact order
$b \sim U(0.4, 2.6)$ \emph{unknown}, so the convergence rate
$n^{-1/b}$ varies by six orders of magnitude across the prior (targets
are trained in $\log(-\Root)$ space). The learned model attains
95.5\%/90.7\% coverage overall with interval-width tracking correlation
0.985, and it stays within three points of nominal on \emph{every}
contact-order slice: 97.3\% on $b \in [0.4, 0.7]$, 96.7\% on $b \in
[0.9, 1.1]$, and 94.0\% on $b \in [2.0, 2.6]$. The standard bootstrap
swings from 96.8\% to 49.0\% across the same slices, and subsampling
from 99.8\% to 20.5\% (Table~\ref{tab:m4c_beta_max}). The network
learns a data-conditional convergence rate; no classical method in this
comparison has a mechanism for doing so.
\begin{table}[htbp]
\centering
\caption{Bounded max with unknown endpoint contact order, b $\sim$ U(0.4, 2.6), n = 200. Nominal coverage is 0.95 (Cov. 95) and 0.90 (Cov. 90); interval lengths and W1 distances to the true root distribution are normalized per dataset by the true root scale. Blank cells: metric not defined for that method.}
\label{tab:m4c_beta_max}
\begin{tabular}{lcccc}
\toprule
Method & Cov. 95 & Cov. 90 & Len. 95 & W1 truth \\
\midrule
standard\_bootstrap & 0.735 & 0.610 & 3.02 & 1.103 \\
subsampling\_m=34 & 0.642 & 0.602 & 3.77 & 1.147 \\
learned\_raw & 0.935 & 0.881 & 5.01 & 0.448 \\
learned\_recal\_(ours) & 0.955 & 0.907 & 6.03 & 0.508 \\
\bottomrule
\end{tabular}
\end{table}

\subsection{Real-data validation: value-at-risk for daily market returns}
\label{sec:realdata}
Real data has no known truth, so we recover one: for each of five
market series (Nasdaq Composite, S\&P 500, EUR/USD, USD/JPY, GBP/USD;
daily losses from FRED, 2{,}511 to 6{,}666 observations each), we treat
the empirical distribution of the \emph{full} series as the population,
so the true VaR$_{0.99}$ is the 99\% quantile of the whole series and is
known exactly. We then draw 2{,}000 i.i.d.\ subsamples of $n = 200$ per
series and apply the NTS-trained specialist completely unchanged, with
frozen weights and frozen recalibration, against the classical methods
on identical subsamples. Real losses are outside the training prior, so
the out-of-family analysis of Section~\ref{sec:ood} sets the
expectation: the Student-$t$ VaR transfer covered 86.1\%.

\begin{figure}[htbp]
\centering
\includegraphics[width=0.72\textwidth]{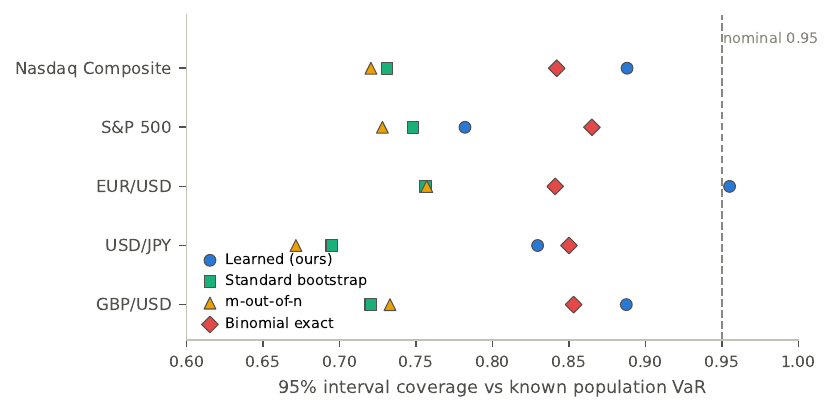}
\caption{Real-data validation. 95\% interval coverage against the known
population VaR$_{0.99}$ of each full return series, over 2{,}000
subsamples of $n = 200$ per series. The learned method (applied
unchanged, out of prior) is closest to nominal on four of five series;
the binomial exact interval cannot exceed 0.866 by construction.}
\label{fig:realvar}
\end{figure}

The results (Figure~\ref{fig:realvar}, Table~\ref{tab:m5_real_var})
land almost exactly there.
Mean 95\% coverage across the five series is 0.868 for the learned
method, against 0.730 for the standard bootstrap, 0.722 for
$m$-out-of-$n$, and 0.850 for the exact order-statistic interval (which
cannot exceed its 0.866 ceiling). The learned method is closest to
nominal on four of five series, reaching 0.955 on EUR/USD. The
exception is the S\&P 500, the shortest series (ten years), where a
handful of 2020 crash days constitute the entire population tail and
every method undercovers; the conservative order-statistic interval is
least bad there. One caveat is stated plainly: i.i.d.\ subsampling
removes serial dependence, so this exercise tests transfer to real
marginal loss distributions, not time-series forecasting.

\begin{table}[htbp]
\centering
\caption{Real-data VaR 0.99: coverage against the known population quantile of each full return series (2000 i.i.d. subsamples of n = 200 per asset; Len. 95 column is median width / true VaR)}
\label{tab:m5_real_var}
\begin{tabular}{lccc}
\toprule
Method & Cov. 95 & Cov. 90 & Len. 95 \\
\midrule
nasdaq / learned\_(ours) & 0.888 & 0.823 & 0.57 \\
nasdaq / standard\_bootstrap & 0.731 & 0.703 & 0.67 \\
nasdaq / m\_of\_n\_m100 & 0.721 & 0.701 & 0.55 \\
nasdaq / binomial\_exact & 0.842 & 0.842 & 0.61 \\
sp500 / learned\_(ours) & 0.782 & 0.722 & 0.53 \\
sp500 / standard\_bootstrap & 0.748 & 0.712 & 0.80 \\
sp500 / m\_of\_n\_m100 & 0.728 & 0.705 & 0.63 \\
sp500 / binomial\_exact & 0.865 & 0.865 & 0.73 \\
eurusd / learned\_(ours) & 0.955 & 0.913 & 0.57 \\
eurusd / standard\_bootstrap & 0.756 & 0.717 & 0.51 \\
eurusd / m\_of\_n\_m100 & 0.757 & 0.724 & 0.44 \\
eurusd / binomial\_exact & 0.841 & 0.841 & 0.44 \\
usdjpy / learned\_(ours) & 0.830 & 0.749 & 0.55 \\
usdjpy / standard\_bootstrap & 0.695 & 0.650 & 0.75 \\
usdjpy / m\_of\_n\_m100 & 0.671 & 0.646 & 0.60 \\
usdjpy / binomial\_exact & 0.850 & 0.850 & 0.67 \\
gbpusd / learned\_(ours) & 0.887 & 0.849 & 0.57 \\
gbpusd / standard\_bootstrap & 0.721 & 0.679 & 0.69 \\
gbpusd / m\_of\_n\_m100 & 0.733 & 0.694 & 0.56 \\
gbpusd / binomial\_exact & 0.853 & 0.853 & 0.63 \\
\bottomrule
\end{tabular}
\end{table}

\subsection{Replication}
All experiments were replicated under two additional training seeds
(re-randomizing training data, validation data, and initialization,
with test sets held fixed for paired comparison). Every gate passed in
every replicate; mean 95\% coverage across seeds is 0.951, 0.951,
0.951, and 0.947 for the four families, with seed-to-seed ranges of
0.2 to 0.4 percentage points and $W_1$ ranges of a few percent
(Table~\ref{tab:replication}).
\begin{table}[htbp]
\centering
\caption{Multi-seed replication: mean [min, max] over three training seeds; test sets identical across seeds.}
\label{tab:replication}
\begin{tabular}{lccc}
\toprule
Family & Cov.\ 95 & Cov.\ 90 & W1 truth \\
\midrule
Uniform max & 0.951 [0.949, 0.952] & 0.902 [0.901, 0.905] & 0.025 [0.021, 0.029] \\
Stable mean & 0.951 [0.951, 0.952] & 0.898 [0.897, 0.900] & 0.372 [0.371, 0.375] \\
Pareto Hill & 0.951 [0.950, 0.952] & 0.903 [0.902, 0.904] & 0.050 [0.049, 0.052] \\
NTS VaR 0.99 & 0.947 [0.945, 0.949] & 0.902 [0.900, 0.903] & 0.126 [0.124, 0.129] \\
Beta max (unknown $b$) & 0.955 [0.955, 0.955] & 0.907 [0.907, 0.907] & 0.508 [0.508, 0.508] \\
\bottomrule
\end{tabular}
\end{table}

\section{Limitations}
\label{sec:limitations}
The method's guarantees are prior-relative: coverage is calibrated
marginally over the training prior, and rate-changing shifts outside
the prior produce confidently wrong intervals. Resampling methods share
this failure mode, but here the scope is declared explicitly by the
prior rather than implicitly by asymptotic assumptions. Sample size is
fixed at $n = 200$; varying-$n$ training via quantile-grid
featurization is future work. NTS training-target centering carries a
Monte Carlo error of about 1\% of the root scale. Our theory is
inherited from proper-scoring and PFN consistency arguments rather than
developed for the root estimand; developing that theory is a natural
next step. Detecting out-of-prior inputs at deployment, whether by
density checks in feature space or conformal fallbacks, remains open.

\section{Conclusion}
Where the bootstrap is inconsistent, the sampling distribution of a
root is still learnable: from simulation alone, with one forward
pass at deployment, at nominal coverage, and near provable optimality
where optimality is computable. The prior over data-generating
processes replaces the silent asymptotic assumptions of resampling with
an explicit, criticizable modeling choice. We view pretrained libraries
of root networks, covering many non-regular statistics behind a single
interface, as the natural continuation; our released software takes the
first step.

\appendix
\section{A cautionary tale: input-blind amortized models}
\label{app:cautionary}

An earlier version of this project trained a set network with a
Gaussian-mixture head to match, for each of several \emph{fixed}
data-generating processes, the Monte Carlo distribution of the centered
statistic, so all training examples from a process shared one target
distribution. The results looked spectacular: Kolmogorov-Smirnov
distances to the truth at the finite-simulation floor, large headline
improvements over the bootstrap. They were an artifact. Because every
training example within a process shared the same target, the
loss-optimal solution was a \emph{constant} function of the input: the
network had memorized the Monte Carlo oracle and ignored the data
entirely. The tell-tale experiment took five minutes: predictions were
unchanged when the inputs were replaced by Gaussian noise, or by all
zeros. Every downstream ``finding'' of that version, from architecture
and feature ablations to generalization failures, was explained by this
single degeneracy.

The present design makes that failure structurally impossible and,
separately, detectable. Structurally: every training example draws
fresh parameters from the prior and is supervised by a single
independent root draw, so no two examples share a target and a constant
predictor is suboptimal whenever the root law varies over the prior.
Detectably: the width-tracking diagnostic of
Section~\ref{sec:diagnostics} is run in every experiment, and a
memorizing model scores approximately zero on width tracking. We
recommend both practices, non-shared targets and a mandatory
width-tracking check, to anyone training amortized inference
models. Marginal evaluation metrics such as average distances to
targets or average calibration \emph{cannot} distinguish a conditional
model from a well-placed constant, and the class of methods that looks
excellent while ignoring its input is larger than it appears.



\begin{thebibliography}{99}
\itemsep0pt

\bibitem[Athreya(1987)]{athreya1987}
Athreya, K.~B. (1987). Bootstrap of the mean in the infinite variance
case. \emph{Annals of Statistics} 15(2), 724--731.

\bibitem[Al~Kadhim et~al.(2024)]{alkadhim2023}
Al~Kadhim, A., Prosper, H.~B., and Prosper, O. (2024). Amortized
simulation-based frequentist inference for tractable and intractable
likelihoods. \emph{Machine Learning: Science and Technology} 5, 015020.

\bibitem[Beran(1987)]{beran1987}
Beran, R. (1987). Prepivoting to reduce level error of confidence sets.
\emph{Biometrika} 74(3), 457--468.

\bibitem[Bertail et~al.(1999)]{bertail1999}
Bertail, P., Politis, D.~N., and Romano, J.~P. (1999). On subsampling
estimators with unknown rate of convergence. \emph{JASA} 94(446),
569--579.

\bibitem[Bickel and Freedman(1981)]{bickel1981}
Bickel, P.~J. and Freedman, D.~A. (1981). Some asymptotic theory for
the bootstrap. \emph{Annals of Statistics} 9(6), 1196--1217.

\bibitem[Bickel et~al.(1997)]{bickel1997}
Bickel, P.~J., G\"otze, F., and van Zwet, W.~R. (1997). Resampling
fewer than $n$ observations: gains, losses, and remedies for losses.
\emph{Statistica Sinica} 7, 1--31.

\bibitem[Bickel and Sakov(2008)]{bickelsakov2008}
Bickel, P.~J. and Sakov, A. (2008). On the choice of $m$ in the $m$ out
of $n$ bootstrap and confidence bounds for extrema. \emph{Statistica
Sinica} 18, 967--985.

\bibitem[Dalmasso et~al.(2021)]{dalmasso2021}
Dalmasso, N., Masserano, L., Zhao, D., Izbicki, R., and Lee, A.~B.
(2021). Likelihood-free frequentist inference. arXiv:2107.03920.

\bibitem[Efron(1979)]{efron1979}
Efron, B. (1979). Bootstrap methods: another look at the jackknife.
\emph{Annals of Statistics} 7(1), 1--26.

\bibitem[Hermans et~al.(2022)]{hermans2022}
Hermans, J., Delaunoy, A., Rozet, F., Wehenkel, A., Begy, V., and
Louppe, G. (2022). A trust crisis in simulation-based inference?
\emph{TMLR}.

\bibitem[Hollmann et~al.(2023)]{hollmann2023tabpfn}
Hollmann, N., M\"uller, S., Eggensperger, K., and Hutter, F. (2023).
TabPFN: A transformer that solves small tabular classification problems
in a second. \emph{ICLR}.

\bibitem[Kuleshov et~al.(2018)]{kuleshov2018}
Kuleshov, V., Fenner, N., and Ermon, S. (2018). Accurate uncertainties
for deep learning using calibrated regression. \emph{ICML}.

\bibitem[Loh(1987)]{loh1987}
Loh, W.-Y. (1987). Calibrating confidence coefficients. \emph{JASA}
82(397), 155--162.

\bibitem[Masserano et~al.(2023)]{masserano2023}
Masserano, L., Dorigo, T., Izbicki, R., Kuusela, M., and Lee, A.~B.
(2023). Simulator-based inference with Waldo. \emph{AISTATS}.

\bibitem[M\"uller et~al.(2022)]{muller2022pfn}
M\"uller, S., Hollmann, N., Arango, S.~P., Grabocka, J., and Hutter, F.
(2022). Transformers can do Bayesian inference. \emph{ICLR}.

\bibitem[Nagler(2023)]{nagler2023}
Nagler, T. (2023). Statistical foundations of prior-data fitted
networks. \emph{ICML}.

\bibitem[Nalisnick and Smyth(2017)]{nalisnick2017}
Nalisnick, E. and Smyth, P. (2017). The amortized bootstrap. \emph{ICML
Workshop on Implicit Models}.

\bibitem[Nie and Ro\v{c}kov\'a(2022)]{nie2022}
Nie, L. and Ro\v{c}kov\'a, V. (2022). Deep bootstrap for Bayesian
inference. arXiv:2205.15374.

\bibitem[Politis and Romano(1994)]{politis1994}
Politis, D.~N. and Romano, J.~P. (1994). Large sample confidence
regions based on subsamples under minimal assumptions. \emph{Annals of
Statistics} 22(4), 2031--2050.

\bibitem[Radev et~al.(2020)]{radev2020bayesflow}
Radev, S.~T., Mertens, U.~K., Voss, A., Ardizzone, L., and K\"othe, U.
(2020). BayesFlow: Learning complex stochastic models with invertible
neural networks. \emph{IEEE TNNLS}.

\bibitem[Sainsbury-Dale et~al.(2024)]{sainsbury2024}
Sainsbury-Dale, M., Zammit-Mangion, A., and Huser, R. (2024).
Likelihood-free parameter estimation with neural Bayes estimators.
\emph{The American Statistician} 78(1), 1--14.

\bibitem[Shin et~al.(2021)]{shin2021neuboots}
Shin, M., Cho, H., Min, H.-s., and Lim, S. (2021). Neural
bootstrapper. \emph{NeurIPS}.

\bibitem[Shin et~al.(2020)]{shin2020gms}
Shin, M., Wang, L., and Liu, J.~S. (2020). Scalable uncertainty
quantification via generative bootstrap sampler. arXiv:2006.00767.

\bibitem[Zammit-Mangion et~al.(2025)]{zammit2025review}
Zammit-Mangion, A., Sainsbury-Dale, M., and Huser, R. (2025). Neural
methods for amortized inference. \emph{Annual Review of Statistics and
Its Application} 12.

\end{thebibliography}
\end{document}